# Rapid measurement of the net charge on a nanoparticle in an optical levitation system

Jin-Chuan Wang[1], Cui-Hong Li[1*], Shao-Chong Zhu[1], Chao-Xiong He[1], Zhi-Ming Chen[1], Zhen-Hai Fu[1], and Hui-Zhu Hu[1,2*]

[1] Research Center for Quantum Sensing, Intelligent Perception Research Institute, Zhejiang Lab, Hangzhou 310000, China

[2] State Key Laboratory of Modern Optical Instrumentation, College of Optical Science and Engineering, Zhejiang University, Hangzhou 310027, China

(E-mail: licuihong@zhejianglab.com, huhuizhu2000@zju.edu.cn)

The accurate measurement of the net charge on a nanoparticle is critical in both the research and application of nanoparticles. Particularly, in the field of precision sensing based on optically levitated nanoparticles, the precise measurement of the net charge on a nanoparticle is prerequisite for stable levitation before electric control process and for the detecting of quantities such as the electric field. However, the accurate measurement of the net charge on a levitated nanoparticle still faces challenges. Here, we proposes a method to measure the net charge accurately through its thermally and harmonically driven motional signals. With the numerical calculation of the dynamic temperature of the optically-levitated nanoparticle, the long-term dynamic net charge on the nanoparticle can be tracked. This method can achieve an accurate measurement with the accuracy of better than 5% at thermal equilibrium pressures of more than 10 mbar and accuracy of ~20% till 3 mbar, where the net charge barely changes spontaneously, and nonlinearities in the levitated oscillator is relatively small. As the net charge is integer discrete, the method provides nearly completely accurate measurements of nanoparticles with few charges. The result can be applied to improve the measurement accuracy for mass and density through iteration and this method provides a way toward non-contact characterization of aerology, space dust, etc.

The net charge is one of the most important characteristic parameters of particles, which is of great significance for the research of the chemical reaction of atmospheric particle[1,2], kinetic dynamics of dust particle[3], and micro scale physical and chemical reaction[4]. In the field of precision measurement based on optical traps, the precise measurement of the net charge affects the sensibility for force related parameters[5,6], the ultimate cooling temperature with electric field feedback control method[7] and the levitation stabilities of particles.

Conventional net charge to measure the net charge are mostly contact measurement, such as the measurement method using Zata[8-10] potential and electrostatic force microscopy (EFM)[11]. The former is usually used in liquid environments, and the latter requires substrate material[12,13].

Currently, the method for measuring net charge in an optical levitation system relies on statistical discrete electrical response of nanoparticles with dynamically controlling the value of net charges[14,15]. Generally, ultraviolet discharge or high voltage ionizing methods are applied to the control of net charge on particles[16]. The former method works good on discharging of initially negatively charged particles[15] and the latter can control the net charge on particles in both direction through ionizing gas molecules inside the vacuum chamber[17]. However, both method that requires frequent charge tuning on particles may lead to unstable levitation. And the high voltage discharge process generally needs to be carried out under a specific pressure range [16,18,19]. In high vacuum, a small quantity of gas in the chamber will make it difficult for high-voltage discharge to generate free charges[20]. What is more, the movement of particles becomes more violent under high vacuum, where particles suffer risks of being pushed out of the optical trap by Coulomb force. In low vacuum, the high gas concentration in the chamber will make the ionized free charges quickly neutralized by the gas molecules, which bring difficulties charge adjusting. In addition, the methods that change the initial net charge may be not desired for some chemical related research fields.

In this paper, we proposed and experimentally demonstrated a direct charge calibration method to quickly measure the net charge of the particle based on the inertia-mass parameter of specific material particles. For a particle trapped in an optical trap driven by the electric field $E = E_0\cos(\omega_{dr}t)$, the number of charges $n_q$ can be derived from the electrical driven motion under force $F_{dr} = n_q q_e E_0 \cos(\omega_{dr}t)$ and its inertia-mass $m$. The main measurement error arises from the mass $m$, which can be calibrated through aerodynamic characteristics of particles, and the error is generally within 10%[17,21]. As the number of net charges is integer discrete, we can error-freely estimate the number of charges for a particle with few charges through rounding by proximity. By calculating the changing temperature of the particle in the optical trap, we realize the measurement of the dynamic net charge of the particle in optical trap from 100-0.5 mbar. Further, By comparing with the measuring results from traditional discrete charge calibration method, we demonstrate that the method we proposed has good performance when the pressure P > 3 mbar. Specifically, the error is less than 5% when the pressure P > 10 mbar. The error is less than 20% when the pressure is reduced to 3 mbar where the system is in a thermal non-equilibrium condition. The error increases significantly at a lower pressure due to the nonlinearities of the oscillations of the levitated particles. The charge measurement method proposed here can help improve calibration accuracy of the particle's mass, density and other parameters through iteration. Moreover, the levitation-based single nanoparticle characterization method can be applied for studying lunar dusts and nano-aerosols, whose physical parameters has growing demands to be precisely measured and monitored[22,23].

## 1. Theoretical principles

The motion equation of a particle with mass $m$ and the number of charge $n_q$ captured in an optical trap can be described by a thermally and harmonically driven damped resonator[17]:

$$\ddot{x}+\Gamma\dot{x}+\omega_0^2 x = \frac{F_{th}(t)}{m} + \frac{n_q q_e E_0 \cos(\omega_{dr}t)}{m} \quad (1)$$

Here, $\Gamma$ is the damping rate, with $F_{th}(t)$ being random collisions with residual air molecules in the chamber. $\omega_0$ is the mechanical eigenfrequency of the oscillator. $\omega_{dr}$ is the frequency of the driving electric field. $q_e$ is the elementary charge. $E_0$ is the amplitude of the electric field. The overall power spectral density (PSD) of a thermally driven mechanical resonator subject to external harmonic force reads [17]:

$$S_x(\omega) = S_{th}(\omega) + S_{dr}(\omega) = \frac{2k_B T\Gamma}{m[(\omega_0^2-\omega^2)+(\Gamma\omega)^2]} + \frac{n_q^2 q_e^2 E_0^2}{2m^2}\frac{\tau sinc[(\omega-\omega_{dr})\tau]}{(\omega_0^2-\omega^2)+(\Gamma\omega)^2} \quad (2)$$

Where $k_B$ is the Boltzmann constant and $T$ is room temperature. $S_{th}(\omega)$ is the thermally driven (PSD), and $S_{dr}(\omega)$ is electrically driven PSD, and $\tau$ is the measurement time.

According to the relationship between the mass of particle, the electric force and PSD signal in the theoretical model of electrical drive of the particle levitated in the optical trap, the relationship between the mass of the particle and the net charge on the particle is obtained [17]:

$$\frac{n_q^2}{m} = \frac{8k_B T R_s}{q_e^2 E_0^2 \tau} \quad (3)$$

Where $R_S$ is the ratio of PSD of electric field driving and thermal driving parts, which can be expressed as $R_s = S_{dr}(\omega_{dr})/S_{th}(\omega_{dr})$.

According to the theory of gas dynamics, in the thermal equilibrium environment, the radius of particles can be expressed as[24]:

$$r = 0.619\frac{9\pi}{\sqrt{2}}\frac{\eta_{air}d^2}{\rho k_B T}\frac{P_{gas}}{\Gamma} \quad (4)$$

Here, $\eta_{air} = 1.86\times10^{-5}$ Pa·s is the air viscosity. $d = 0.36\times10^{-9}$ m is the diameter of air molecular. $P_{gas}$ is the ambient pressure. $\rho$ is the particle's density. $\Gamma$ can be obtained by fitting $S(\omega)$ with the least square method. With the radius of the particle $r$, we can obtain the particle mass $m = 4\pi r^3 \rho/3$. Finally, the formula for rapid measurement of the net charge on a nanoparticle based on an optically levitated an optically system reads:

$$n_q = \sqrt{\frac{8k_B T \Gamma R_s m}{q_e^2 E_0^2 \tau}} \quad (5)$$

## 2. Experimental setup

The experimental setup is shown schematically in Fig. 1. A $\lambda=1064$ nm laser beam (YDFL-064-SF-10-CW) is focused by an objective (NA=0.8) to trap a silica nanoparticle at its focus, which is placed inside a vacuum chamber. Nanoparticles suspended in isopropyl alcohol solution are sprayed into the vacuum chamber with a nebulizer, and the optical trap can capture the particles falling nearby. We define the beam propagation direction as Z-axis, while the X-axis and Y-axis correspond to the directions parallel and perpendicular to the polarization direction of the trapping beam in the focal plane. The oscillations of the particle along the three axes are monitored with a four-quadrant photodetector (QPD). A pair of parallel plate electrodes is applied to generate a uniform electric field surrounding the optical trap. The purple glow on the side of the chamber is emitted by a wire connected to a high voltage DC source and is used to control the net charge of the particle via ionizing gas molecules inside the vacuum chamber[5,14].

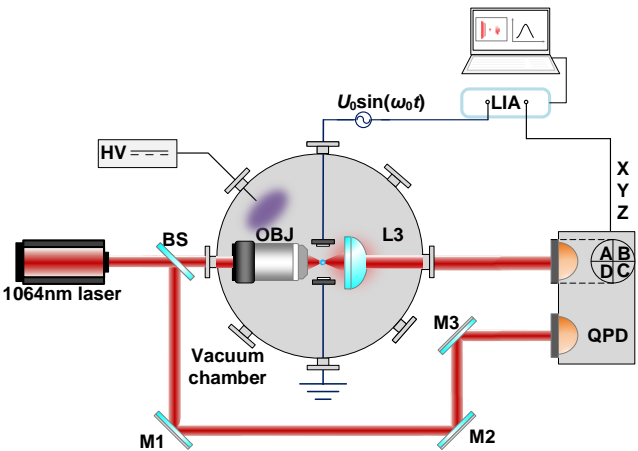

FIG.1 Experimental set-up. A microscope objective (OBJ) focuses a laser beam inside a vacuum chamber, where a single silica nanoparticle is trapped in the focus. The light scattered by the particle is collected with an aspheric lens (AL) and the motion of the particle is detected in a split detection scheme. The signal of the function generator driving the capacitor serves as a reference to a lock-in amplifier (LIA), creating an electric field that drives the charged particle. The purple glow on the side of the chamber is emitted by a wire connected to a high voltage (HV) DC source and is used to control the net charge of the particle. All detector signals are sent to the computer system. BS, beam splitter. QPD, Four-quadrant photodetector.

## 3. Results and discussions

### 3.1 Direct calculation result

To calculate the number of charges the particle carries, the thermally and harmonically driven PSD of the particle along the X-axis is measured under the initial net charge at $P = 5$ mbar. As is shown in Fig.2, the resonance frequency is approximately 153 kHz. Damping rate $\Gamma$, the ratio $R_s$, and particle's radius $r$ is deduced from the PSD, and consequently, $r = 74 \pm 3$ nm, $\Gamma = 4.47 \pm 0.01$ kHz, $R_s = S_{dr}(\omega_{dr})/S_{th}(\omega) = 53.4 \pm 0.9$. When $\tau = 0.015$ s, the net charge number $n_q$ carried by the particle is calculated as $2.37 \pm 0.12$.

Theoretically, the error of our calculation results is not large, and a relatively accurate net charge that the particle carries can be obtained via calculation (see the supplementary materials for specific error characterization). Therefore, we take the number of charges as 2 by proximity rounding. Further, to verify the accuracy of the direct calculation results, we calibrated the net charge of the particle through traditional discrete calibration method.

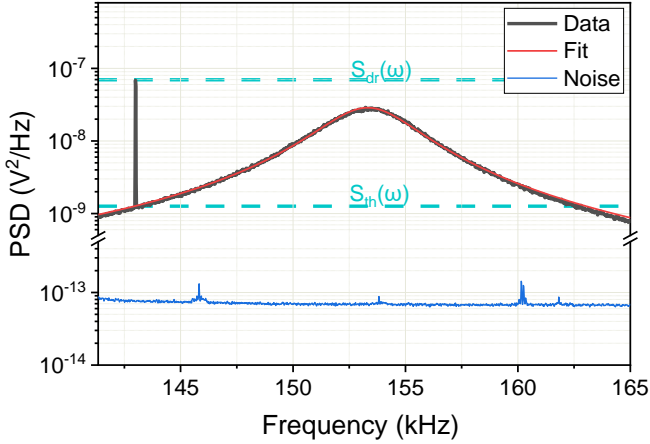

FIG. 2 PSD and Lorentz fitting. PSD(black) of the thermally and harmonically driven oscillator at P = 5 mbar. The resonance frequency is approximately 153 kHz. A driving signal with an amplitude of 22.5 V and the frequency of 143 kHz was applied to the plate electrodes. We fit the thermally driven oscillator with Lorentzian function(red). $R_s = S_{dr}(\omega_{dr})/S_{th}(\omega_{dr}) = 53.4$. The dark blue spectrum at the bottom is the measurement noise.

*3.2 Discrete calibration result*

In Fig.3, we show the measurement results of the net charge on the nanoparticle placed within the optical trap using the discrete charge calibration method. An AC signal $U(t)$ with frequency of 143 kHz is added on the plate electrodes to drive the particle motion. The net charge on the particle is controlled by ionizing gas molecules inside the vacuum chamber at a pressure of 5 mbar. Fig.3 shows the amplitude and phase that the lock-in amplifier(LIA) demodulated from the motion signal in X-axis. When the high-voltage is turned on, the amplitude changes in the form of discrete steps, which is due to the absorption and release of free charges by the nanoparticle. It can be observed intuitively that the amplitude of the first step is 1.4 mV, and the minimum difference between discrete amplitudes is 0.7 mV, which is caused by absorbing or releasing a single elementary charge. Therefore, the initial number of charges on the particle is 2. We can find that the direct calibration result is consistent with that measured by the discrete charge calibration method, with an error of 18.5%. The error between the calculation results and the calibration is greater than the previously estimated error of 5.12% at thermal equilibrium condition(see supplementary information). The main reason is that when the pressure drops to 5 mbar, the nonlinear effect emerges, which increases the error of the fitting results. Furthermore, the error of simulated temperature also affects the results(see supplementary information). Fig.3(b) shows the phase that demodulated by the lock-in amplifier. Clearly, the charge's polarity of the particles remains constant during the controlling process.

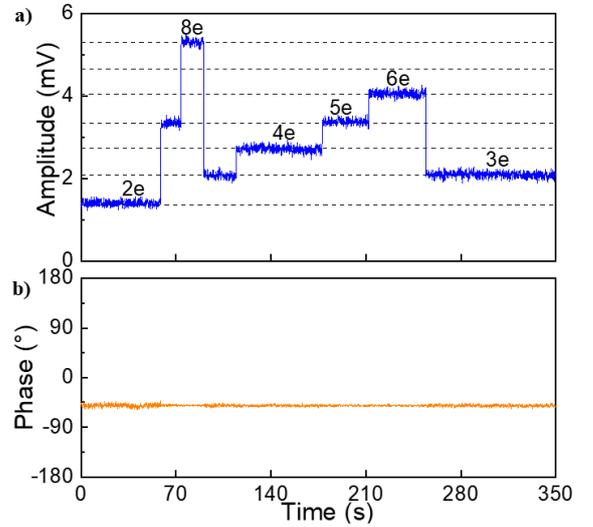

FIG.3. Control the net charge at 5mbar. (a)Amplitude signal demodulated by the lock-in amplifier. The amplitude signal changes in discrete steps after each discharge. (b) Phase signal demodulated by the lock-in amplifier, which reflects that the charge's polarity stays constant over the measurement.

*3.3 discussion*

To test the applicability of this method, the number of charges on the particle at different pressures is measured with both methods. The pressure is reduced from 100 mbar to 0.5 mbar and five air pressure values are chosen, respectively 100 mbar, 30 mbar, 10 mbar, 3 mbar, 1 mbar and 0.5 mbar. Firstly, the measurement is conducted with reduction of pressure from 100 mbar to 0.5 mbar. After going through high vacuum, the pressure is back increased from low pressure, and the charge measurements are conducted for a second time from 0.5 mbar to 100 mbar. The charge measurement results during the pressure reduction process and increasing process are respectively shown in Fig. 4.(a) and (b).

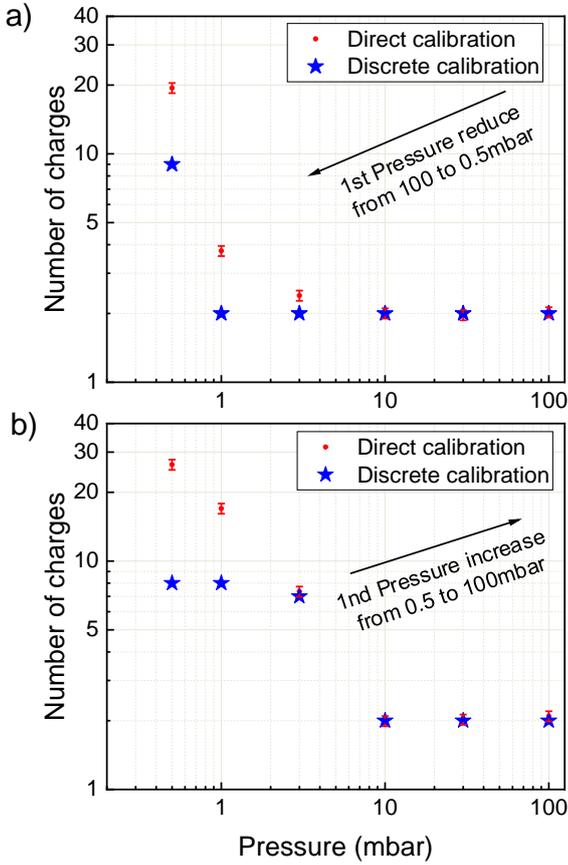

FIG.4 Contrast diagram of charge calculation results and calibration results at different pressures. (a) the contrast diagram between the calculation result and the calibration result when the pressure decrease from 100 mbar to 0.5 mbar. (b) The contrast diagram between the calculation result and the calibration result when the pressure rises from 0.5 mbar to 100 mbar. The error increases at 3 mbar. When the pressure drops to approximately 1 mbar, the error significantly increases.

It can be observed that the direct calculation results are basically constant and consistent with the discrete calibration results when the pressure $P > 3$ mbar. The error between the two results are within 5% at 100, 30 and 10 mbar. The error increases at 3 mbar, and errors significantly increase at pressures lower than 1 mbar. It is obvious that the charge of the trapped nanoparticle can be error-freely estimated by its calculated result at pressures of more than 3 mbar.

Fig.5 shows the diagram of the change of the phase-locked signal's amplitude with reducing the pressure from 100 mbar to 0.5 mbar with the frequency and amplitude of the driving electric field unchanged. The amplitudes changed with the pressure may arise from two reasons. For one thing, the net charge may slightly change during the process of changing the pressure. For another, the response to the driving electric field is dependent on the pressure. The latter is dominant, which states the discrete calibration method has the shortage that the discrete amplitude step needs calibrating at each pressure. And there is a high risk of particles falling from the optical trap during this process. In addition, when the pressure drops to less than 1 mbar, the amplitude of the phase-locked signals seems to vary irregularly with the pressure, which brings extra uncertainty to the discrete calibration method. The reason for the increased charge measurement error and the strong amplitude hopping phenomenon at high vacuums may arise from the influence of nonlinear effects, which is caused by the fact that the motions of particles become more violent and the instability of particle motions are enhanced when the pressure decreases[25].

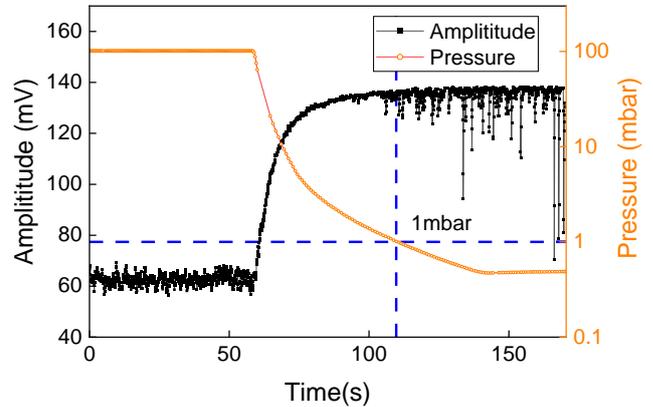

FIG.5 Change of oscillation amplitude of the particle in response to a driving voltage with decreasing pressure. When the air pressure decreases, the oscillation signal amplitude of the particle rises sharply. After reaching 1mbar, the signal shows a strong hop phenomenon.

For an optical trap constructed by a Gaussian beam, its potential deviates from quadratic shape, which induces a nonlinear natural frequency shift dependent on the motion of trapped nanoparticles. The trapped particles near the equilibrium position in the optical trap have a fixed eigenfrequency $\omega_0$ when the amplitude is far less than the beam waist and the optical potential can be perfectly approximated by a harmonic potential. However, the optical potential will become an-harmonic when the amplitude of the oscillator increases[25]. Fig.6 shows the thermally driven PSD of the particle at pressures of 3 mbar, 1 mbar and 0.05 mbar. Time trace of the particle motion is divided into long time trace(blue) and short time trace (red and black). At the pressure of 3 mbar, the long time trace almost coincides with the short time traces. When the pressure drops to 1 mbar, the frequency of the curve fluctuates obviously. And when the pressure reaches 0.05 mbar, the fluctuation becomes more severe. The frequency fluctuation caused by the nonlinear effect will brings errors to the fitting calculation results of damping rate $\Gamma$ and $S_{th}(\omega)$, and consequently, this error will propagate to the results of the result of particle radius r, mass m and the number of charges $n_q$. Furthermore, in the process of reducing the pressure, we also found that the number of charges $n_q$ is basically constant before the pressure drops to 3 mbar. The number of charges changes mainly below 1 mbar, which may be related to the chemical change process of the particle in the optical trap[26].

The traditional way to suppress the nonlinear effect of particle motion in high vacuum is cooling the center of mass motion of nanoparticle through feedback systems[27]. Unfortunately, when measuring the net charge of particle, the electric field driving motion of particles are inevitably cooling-controlled, bringing errors to the net charge measurement. Although, two axis cooling of the trapped particles on two axes can suppress the non-linear fluctuation motion of nanoparticle along the third axis without influencing its electrical-driving part, the cooling effect cannot support precise charge measurement. (see supplementary material II).

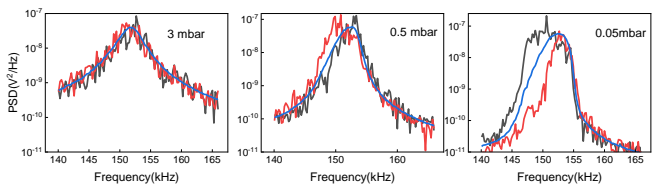

FIG.6 Nonlinearity-induced frequency fluctuations. The red line and black line are the short time traces, while the blue line is the long time trace. With the decrease of the pressure, the nonlinear effect is significantly enhanced.

## 4. Conclusions

In conclusion, we propose and demonstrate a simple method to rapidly measure the net charge on a particle based on its inertial mass parameters and harmonically and thermally driven PSD signals. This method can quickly measure the net charge on a particle within a wide pressure range, and there is no risk of losing particles from the optical trap. This method does not require manipulating of net charge, with no changing to the initial charge amount. By introducing temperature simulation, the dynamic net charge can be tracked. The direct calibration results at the pressure of P > 3 mbar is precise, with error of less than 5% when P ≥ 10 mbar and less than 20% at 3 mbar. Combining the charge integer characteristic, we can realize fast error-free estimation for particles with low charge amounts. When the pressure is lower, due to the significant enhancement of nonlinearity, charge estimation error arises. In addition, this method can also be applied to the measurements in other levitation systems, such as Paul traps, magnetic traps[28,29], etc. Through the measurement of net charge, more precise calibration of mass and density can be achieved iteratively[17,30], and the method can be applied to in situ non-contact characterization of other suspension nanoparticles such as dusts and small aerosols[31-33].

**SUPPLEMENTARY MATERIAL**

For details on error propagation and cooling, see the supplementary materials Error propagation, Error correction at 5 mbar and The effect of cooling on direct charge measurement.

**ACKNOWLEDGMENTS**

This research was funded by the National Natural Science Foundation of China (42004154) and Major Scientific Project of Zhejiang Laboratory (2019MB0AD01, 2020MB0AL03).

## SUPPLEMENTARY MATERIALS

### I. Error propagation

In order to determine the error in the calculation, a careful study of all the sources of errors has to be carried out based on the existing parameter conditions. For several variables and constants, we can neglect the corresponding uncertainty. $q_e$, $k_B$ and $\tau$. are all constants, and $R_s$ is calculated from the measured $S_{dr}(\omega)$ and $S_{th}(\omega)$, which are only affected by statistical errors. The error(thermal equilibrium) of mass and the number of charges in equations (4) and (5) is given by:

$$\frac{\sigma_m}{m} = \sqrt{(2\frac{\sigma_\rho}{\rho})^2 + (3\frac{\sigma_T}{T})^3 + (3\frac{\sigma_\Gamma}{\Gamma})^3 + (3\frac{\sigma_{P_{gas}}}{P_{gas}})^3} \quad (1)$$

$$\frac{\sigma_{n_q}}{n_q} = \sqrt{(\frac{\sigma_{E_0}}{E_0})^2 + (\frac{1}{2}\frac{\sigma_T}{T})^2 + (\frac{1}{2}\frac{\sigma_\Gamma}{\Gamma})^2 + (\frac{1}{2}\frac{\sigma_{R_s}}{R_s})^2} \quad (2)$$

The uncertainty of various parameters that cause errors in thermal equilibrium in the experiment is shown in the following table:

Table I Uncertainties table. Black font indicates errors that cannot be ignored. Gray font indicates that $\sigma_{z_i} \sim 0$.

| Quantity | Value $z_i$ | Error $\sigma_{z_i}/z_i$ |
|---|---|---|
| $T$ | 300 K | 5‰[a] |
| $S_{th}(\omega)$ | 1.29×10$^{-8}$ V$^2$/Hz | 1.4 ‰[c] |
| $S_{dr}(\omega)$ | 6.89×10$^{-8}$ V$^2$/Hz | 1.3%[c] |
| $\eta_{air}$ | 1.82×10$^{-5}$ Pa·s | 0.03‰[1] |
| $d$ | 0.353×10$^{-9}$ m | —[2] |
| $P_{gas}$ | 100~0.133 mbar | 2‰[d] |
| $E_0$ | 3800 V/m | 1.1%[3] |
| $\rho$ | 2010 kg/m$^3$ | 4.9%[4] |
| $k_B$ | 1.380×10$^{-23}$ J/K$^{-1}$ | 5.72×10$^{-7}$[5] |
| $q_e$ | 1.602×10$^{-19}$ C | 6.1×10$^{-9}$[5] |
| $\tau$ | 0.015 s | 1 ppm[e] |
| $m$ | 3.406 fg | 9.97% |
| $n_q$ | 2.37 | 5.12% |

[a] This error comes from temperature measurement error.
[b] This error comes from the error brought by the fitting process.
[c] This error comes from the measurement error caused by data fluctuation during data acquisition.
[d] This error comes from the measurement error of vacuum gauge.
[e] Nominal value from the datasheet of lock-in amplifier (Zurich Instruments MFLI).

### II. Error correction at 5 mbar

In an optical levitation system, thermal non-equilibrium effect must be considered at low air pressure. Fig.7 shows the internal temperature of the trapped silica nanoparticle calculated for varying gas pressure considering laser absorption, thermal conduction and black body radiation[3].

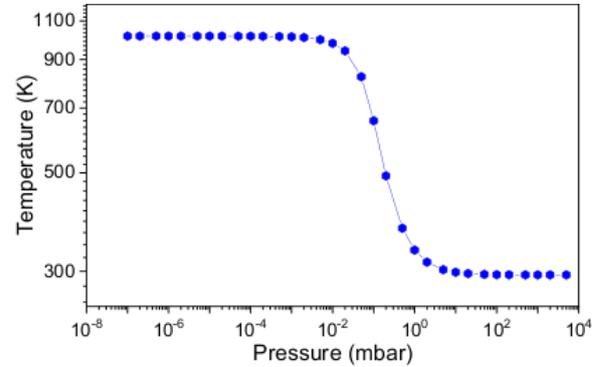

FIG.7 The calculated internal temperature of the trapped nanoparticle with pressure ranging from 104 mbar to 10-7 mbar.

In our system, the calculated internal temperature of the trapped particle at 5 mbar is 310 K. However, the COM (center of mass) temperature of the system inevitably differs from the calculated result. In addition, the inconspicuous nonlinear effect of the trapping system also may leads to the increase of thermal motion related measurement error.

### III. The effect of cooling on direct charge measurement

In order to suppress the nonlinear effect of particle motion in high vacuum, feedback cooling of the center of mass motion of trapped nanoparticle is conducted[6]. Different cooling conditions are tested. We found that triaxial cooling can suppress drift and narrow linewidth. However, the electric field applied on X-axis are inevitably cooling-controlled when measuring the net charge. Two-axis cooling along Y and Z axis can only bring X-axis cooling to a small extent, which is not enough to support the research needs. Fig.7 compares the motion stability of the trapped nanoparticles at 3 mbar, 0.5 mbar and 0.05 mbar without feedback cooling and with feedback cooling along Y and Z axis. It can be observed that the effect of the center of mass motion cooling of the nanoparticle in two axis on restraining the nonlinearity of particle motion is not adequate.

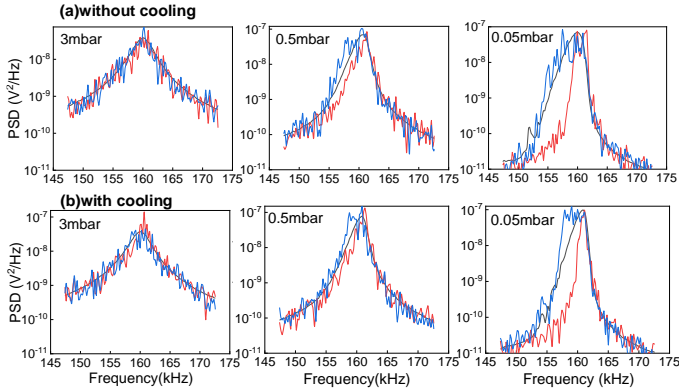

FIG.8 Suppression the nonlinearity motion on X axis through feedback cooling. The red line and black lines are the short time traces, while the blue line is the long time trace. With the decreasing of the pressure, the nonlinear effect significantly enhances. (a) the Nonlinearity-induced frequency fluctuations on X-axis without cooling. (b) the Nonlinearity-induced frequency fluctuations on X-axis with cooling along Y and Z axis.